%% file: ANA-STDM-2024-01-PAPER.tex
\documentclass[cernpreprint, atlasdraft=false, USenglish, texmf, orcidlogo]{atlasdoc}
\usepackage{atlaspackage}
\usepackage{atlasbiblatex}

\usepackage{atlasphysics}
\graphicspath{{logos/}{figures/}}

\usepackage{ANA-STDM-2024-01-PAPER-defs}
\usepackage{multirow}

\AtlasTitle{Observation of pair production of longitudinally polarized $Z$ bosons in four-lepton final state with the ATLAS detector}

\AtlasVersion{2.0}

\AtlasAbstract{%
The production of two longitudinally polarized $Z$ bosons is measured using
proton–proton collision data recorded by the ATLAS experiment at a center-of-mass energy $\sqrt{s}=13.6$~TeV
and with an integrated luminosity of 164~fb$^{-1}$.
The $ZZ$ candidate events are reconstructed using two same-flavor opposite-charge electron or muon pairs.
The polarization fractions of the $Z$ boson pair are measured inclusively in the fiducial phase space and also in two
regions of the four-lepton invariant mass, and the results are consistent with the Standard Model predictions.
After a statistical combination with a previous measurement at $\sqrt{s}=13$~TeV,
the first observation of the production of two simultaneously longitudinally polarized $Z$ bosons is obtained,
with observed and expected significances of \sigObsmuZZLLComb\ and \sigExpmuZZLLComb\ standard deviations, respectively.
}

\AtlasRefCode{STDM-2024-01}

\PreprintIdNumber{CERN-EP-2026-191}

\AtlasJournal{PRL}
\AtlasCoverEgroupAnalysisTeam{atlas-stdm-2024-01-analysis-team@cern.ch}

\hypersetup{pdftitle={ATLAS document},pdfauthor={The ATLAS Collaboration}}

\begin{document}

\maketitle

\clearpage

%
%

After the discovery of the Higgs boson~\cite{HIGG-2012-27,CMS-HIG-12-028}, the scrutiny of Electroweak Symmetry Breaking (EWSB) has become a primary focus of research at the Large Hadron Collider (LHC).
The polarization measurement of massive weak bosons provides a sensitive probe of the EWSB, which generates the masses of the $W$ and $Z$ bosons and gives rise to their longitudinal polarization states.

Precise polarization measurements have been performed with LHC data with single bosons such as in $W$ boson production~\cite{CMS-EWK-10-014, STDM-2011-37, STDM-2020-07},
$Z$ boson production~\cite{STDM-2014-10,CMS-SMP-13-010,CMS-SMP-23-007}, and $W$ bosons from top-quark decays~\cite{TOPQ-2016-02,CMS-TOP-13-008,TOPQ-2018-02}.
Single-boson polarization states have also been measured in $WZ$ production by the ATLAS~\cite{STDM-2018-03} and CMS~\cite{CMS-SMP-20-014} Collaborations.
The increasing amount of data delivered by the LHC has enabled experimental studies of the joint-polarization states of pairs of weak bosons.
ATLAS has measured the joint-polarization states of longitudinally or transversely polarized bosons in inclusive $WZ$ production~\cite{STDM-2022-01}, where the fraction of diboson events with a simultaneous longitudinal polarization~(LL) was observed with a significance of $7.1$ standard deviations.
In the case of two $Z$ bosons, ATLAS has reported evidence of two longitudinally polarized $Z$ bosons~(\ZLZL) with observed significance of $4.3$ standard deviations~\cite{STDM-2021-05}.
Both ATLAS~\cite{STDM-2019-29} and CMS ~\cite{CMS-SMP-20-006} have also measured the production of polarized $W$ bosons in EW production of the same-sign $W^{\pm}W^{\pm}$ boson pairs, and ATLAS has reported evidence of at least one longitudinally polarized $W$ boson~\cite{STDM-2019-29}.

This Letter presents the first observation of the production of \ZLZL\ in the decay channel $ZZ\to \fourl$,
where $\ell$ and $\ell^{\prime}$ can be an electron or a muon.
$Z$ boson candidates are reconstructed with same-flavor, opposite-charge (SFOC) electron or muon pairs.
Events are required to have two $Z$ boson candidates and are split into two signal regions (SR) based on the invariant mass of the four-lepton system (\mfl),
to exploit the kinematic differences between different polarization components.
The $Z$ boson polarization is defined in the center-of-mass frame of the two $Z$ bosons.
For the dominant quark-initiated $ZZ$ production, the longitudinal contribution is strongly suppressed relative to the transverse mode at high \mfl.
In the high-energy limit, the Goldstone Boson Equivalence Theorem~\cite{PhysRevD.10.1145,Vayonakis:1976vz,CHANOWITZ1985379} together with the predominantly light-quark initial states leads to a rapidly decreasing LL fraction.
Therefore, measuring the dependence of the LL component as a function of the diboson invariant mass
provides a crucial test of the EWSB mechanism and a sensitive probe of possible deviations from the SM.

The data recorded by ATLAS during Run~3 of the LHC corresponding to an integrated luminosity of 164~fb$^{-1}$ at a center-of-mass energy $\sqrt{s}=13.6$~TeV is used. A statistical combination with the corresponding measurement using Run~2 data at $\sqrt{s}=13$~TeV~\cite{STDM-2021-05} is performed.

%


%

%

%
\newcommand{\AtlasCoordFootnote}{%
ATLAS uses a right-handed coordinate system with its origin at the nominal interaction point (IP)
in the center of the detector and the \(z\)-axis along the beam pipe.
The \(x\)-axis points from the IP to the center of the LHC ring,
and the \(y\)-axis points upwards.
Polar coordinates \((r,\phi)\) are used in the transverse plane,
\(\phi\) being the azimuthal angle around the \(z\)-axis.
The pseudorapidity is defined in terms of the polar angle \(\theta\) as \(\eta = -\ln \tan(\theta/2)\) and is equal to the rapidity
$ y = \frac{1}{2} \ln \left( \frac{E + p_z}{E - p_z} \right) $ in the relativistic limit.
Angular distance is measured in units of \(\Delta R \equiv \sqrt{(\Delta y)^{2} + (\Delta\phi)^{2}}\).
}

The ATLAS experiment~\cite{PERF-2007-01,GENR-2019-02} at the LHC is a multipurpose particle detector
with a forward--backward symmetric cylindrical geometry and a near \(4\pi\) coverage in
solid angle.\footnote{\AtlasCoordFootnote}
It consists of an inner tracking detector~(ID) surrounded by a thin superconducting solenoid, electromagnetic~(EM) and hadronic calorimeters,
and a muon spectrometer~(MS) based on three large superconducting air-core toroidal magnets.
A software suite~\cite{SOFT-2022-02} is used in data simulation, in the reconstruction
and analysis of real and simulated data, in detector operations, and in the trigger and data acquisition
systems of the experiment.

Simulated Monte Carlo (MC) samples are used to model the signal and background processes, optimize the event selection, and construct the templates
to extract the polarization fractions of the two $Z$ bosons.
All simulated MC events are passed through a full simulation of the ATLAS detector response~\cite{SOFT-2010-01} using \GEANT~\cite{Agostinelli:2002hh}.
The effect of multiple interactions in the same and neighbouring bunch crossings (\pileup) was included in the simulation to match data~\cite{SIMU-2020-01}.
The next-to-next-to-leading-order (NNLO) parton distribution function (PDF) set of \NNPDF[3.0]~\cite{Ball:2014uwa} was used in the matrix-element~(ME)
calculation for all processes.
The inclusive $ZZ$ production was generated with
\progname{Sherpa}~2.2.16~\cite{Gleisberg:2008ta,Gleisberg:2008fv,Cascioli:2011va}, with ME calculation accuracy at next-to-leading-order~(NLO)
in the strong coupling $\alpha_s$ for final states with up to one additional parton, and at leading-order~(LO) accuracy for up to three additional partons.
The ME calculations utilized \progname{OpenLoops}~\cite{Cascioli:2011va} for the virtual QCD corrections, and
were merged with the \progname{Sherpa} parton shower (PS) using the ME+PS@NLO prescription~\cite{Hoeche:2012yf}.
This sample is referred to as \qqtoZZ.
Higher-order contributions from \ggtoZZ\ production~($\mathcal{O}(\alpha_s^2\alpha_{\mathrm{EW}}^2)$) and
EW $ZZjj$ production~($\mathcal{O}(\alpha_{\mathrm{EW}}^4)$), where $\alpha_{\mathrm{EW}}$ is the EW coupling,
were simulated with \progname{Sherpa}~2.2.12 and
\progname{Sherpa}~2.2.14, respectively, at LO precision with up to one extra parton, with the same PS modeling as described above.
The \ggtoZZ~sample includes the box-diagram \ggtoZZ~process and the off-shell Higgs boson production with $H^{*} \rightarrow ZZ$ decays.

Dedicated samples for specific polarization states, \ZTZT, \ZTZL\ and \ZLZL,
were simulated using \MGNLO[3.5.5]~\cite{Alwall:2014hca, BuarqueFranzosi:2019boy} at LO in $\alpha_s$
and interfaced to \PYTHIA[8.312]~\cite{Sjostrand:2014zea} for PS and hadronization,
following the same approach as in Ref.~\cite{STDM-2021-05}.
For \qqtoZZ\ production, samples were generated at LO accuracy for up to two extra partons in the final state,
using the CKKW-L scheme~\cite{Lonnblad:2001iq, Lonnblad:2011xx}.
For EW $ZZjj$ production, samples were produced at LO accuracy.
Alternative polarized samples for \qqtoZZ\ were also simulated at the particle level using \progname{Sherpa}~3
at approximate NLO accuracy for zero and one partons in the final state~\cite{Hoppe:2023uux} and using the ME+PS@NLO merging procedure.
The narrow-width approximation was used and all polarization contributions were computed and stored as alternative event weights.
The interference contribution is obtained by the subtraction of event weights between the inclusive ME weight and the sum of the three polarized ME weights.

For polarized \ggtoZZ\ production, no MC generators capable of directly producing polarized events are currently available,
and fixed-order predictions exist only at LO~\cite{Carrivale:2025mjy}.
In the Run~2 analysis, a dedicated reweighting method was used to obtain polarized \ggtoZZ\ templates.
However, significant discrepancies were recently observed when comparing the reweighted predictions with fixed-order calculations~\cite{Carrivale:2025mjy}.
Therefore, the inclusive \ggtoZZ\ process is treated as a background in this analysis to be more conservative.
Additional samples were generated to estimate the background contribution from events with at least four prompt leptons,
arising from $t\bar{t}Z$ and tri-boson ($VVZ, V = W, Z$) production.
The $t\bar{t}Z$ process was simulated with \MGNLO[2.7.3]+\PYTHIA[8.230].
The $VVZ$ process was simulated using \progname{Sherpa}~2.2.16, with a similar configuration as with the \progname{Sherpa} \qqtoZZ\ sample.
%
%
%


The data used were recorded with single-lepton triggers, with an overall efficiency of approximately 100\% for events selected in the SR.
The object and event selection closely follows the previous $ZZ$ polarization analysis at $\sqrt{s}=13$~TeV~\cite{STDM-2021-05}.
All events must have at least one primary vertex~\cite{ATL-PHYS-PUB-2019-015} with at least two associated ID tracks.
Electrons are reconstructed from topological clusters of EM calorimeter cells matched to ID tracks~\cite{EGAM-2018-01, EGAM-2021-01, EGAM-2021-02}.
Electron candidates are required to satisfy $\pT > 7$~GeV, $|\eta| < 2.47 $,
and must be matched to the primary vertex by having a longitudinal impact parameter $|z_{0}\sin\theta|<0.5$~mm, where $\theta$ is the polar angle of the track.
\textit{Baseline} electrons are defined using the `VeryLooseLH' identification criteria~\cite{EGAM-2018-01, EGAM-2021-01} and are used for the non-prompt background estimate.
\textit{Signal} electrons are used in the final event selection and are required to satisfy tighter requirements,
including the `LooseAndBLayerLH' identification criteria~\cite{PERF-2017-01},
the `Loose\_VarRad' isolation criteria~\cite{EGAM-2021-01},
and transverse impact parameter significance of $|d_{0}|/\sigma_{d_{0}} <  5 $, where $d_0$ is the transverse impact parameter relative to the beamline, and $\sigma_{d_{0}}$ is its uncertainty.
Muon candidates are reconstructed by combining the tracks in the ID and MS~\cite{MUON-2018-03, MUON-2022-01}.
Both \textit{baseline} and \textit{signal} muon candidates must satisfy the `Loose' identification criteria~\cite{MUON-2018-03},
and have $\pT>5$~GeV, $|\eta|<2.5$, and $|z_{0}\sin\theta|<0.5$~mm.
\textit{Signal} muons are further required to have $|d_{0}|/\sigma_{d_{0}} <  3 $, and satisfy the `PflowLoose\_VarRad' isolation criteria~\cite{MUON-2018-03}.
Jets are reconstructed from particle-flow objects~\cite{PERF-2015-09} using the anti-$k_t$ algorithm~\cite{Cacciari:2008gp,Fastjet}
with radius $R=0.4$.
Candidate jets are required to satisfy $\pT > 30$~GeV and $|\eta|<4.5$, and are used in the non-prompt background estimate studies.

Events are required to have at least four \textit{baseline} leptons, with the two highest-\pT leptons satisfying $\pT > 27$ and 20~GeV, respectively.
In each event, all possible SFOC lepton pairs are constructed with the requirement of the invariant mass $m_{\ell \ell} > 5$~GeV and $\Delta R_{\ell\ell} > 0.05$ to suppress
low-mass resonance contributions.
A quadruplet is then formed from two SFOC pairs with the smallest sum of $|m_{\ell\ell} - \mZ|$, where \mZ\ is the $Z$ boson pole mass.
The SFOC lepton pair with the largest dilepton absolute rapidity, $|y_{\ell\ell}|$, is defined as the leading $Z$ boson candidate.
To ensure the selection of on-shell $ZZ$ events, the invariant masses of both SFOC lepton pairs are required to be $|m_{\ell\ell}- \mZ| < 10 $~GeV and
the invariant mass of the quadruplet must satisfy $m_{4\ell} > 180$~GeV.
Each lepton of the quadruplet is then required to satisfy the \textit{signal} lepton definition for the SR selection.
The SR is further split into two categories, the low-mass SR with $180~\GeV < m_{4\ell} < 240~\GeV$, and the high-mass SR with $m_{4\ell} > 240$~GeV.


%
Two sources of non-$ZZ$ background are considered and have a combined contribution of less than 2\% of the total yield.
The irreducible background consisting of prompt leptons from $t\bar{t}Z$ and fully leptonic decays of $VVZ$ processes are estimated by using MC simulations.
The reducible background is estimated with a data-driven \textit{fake-factor} method, following the methodology described in Refs.~\cite{STDM-2021-05,STDM-2022-17}.
This contribution mainly arises from the production of $Z$+jets, \ttbar, and $WZ$, with non-prompt leptons originating from hadron decays or misidentification of jets.
A non-prompt enriched control region~(CR) is defined by selecting events in which one or more leptons in the quadruplet fails the signal lepton definition.
The prompt backgrounds are subtracted from the data yield in this CR to estimate the non-prompt yield and extrapolated to the SR using a fake-factor,
which is estimated from data in dedicated $Z$+jets and \ttbar\ enhanced control regions.


%
A boosted decision tree~(BDT)~\cite{TMVA} is trained to separate the production of \ZLZL\
from \ZTZX\ (\ZTZT\ or \ZTZL).
Polarization sensitive angular observables are used as input variables for the BDT:
\dphio\ and \dphit, the azimuthal angles, defined in the four-lepton center-of-mass frame, between the leptons forming the leading and sub-leading $Z$ boson candidates, respectively;
\costho\ and \costht, where \tho(\tht) is the angle between the negatively charged final-state lepton in the \Zo(\Zt) rest frame and the direction of flight of the \Zo(\Zt) boson in the four-lepton rest frame;
and
\costhst, where \thst\ is the production angle of the \Zo\ defined in the four-lepton rest frame.
The \MGNLO\ polarized MC samples for \qqtoZZ\ and EW $ZZjj$ production are used in the BDT training.
To account for higher-order corrections, an event-by-event reweighting approach developed for the Run~2 analysis~\cite{STDM-2021-05, ATL-PHYS-PUB-2025-021}
is used to obtain polarized templates.
The procedure employs a sequence of reweighting steps based on either one-dimensional or two-dimensional~(2D) observables.
The combined NLO QCD and EW corrections calculated at fixed-order with the \MoCaNLO\ program~\cite{Denner:2021csi,Denner:2026phn} for each polarization state, parameterized as a function of \costho, are applied to the corresponding polarized \MGNLO\ MC samples.
The interference template is obtained by reweighting the inclusive \SHERPA[2] MC sample as a function of the BDT score, using weights derived from the particle-level \SHERPA[3] prediction.
A closure test is performed by comparing the sum of the resulting polarized and interference templates with the reference inclusive \SHERPA[2] prediction.
To account for residual non-closure effects, an additional 2D reweighting as a function of \costhst and \dphio is applied to the three polarization templates.
The final \ZTZT, \ZTZL, \ZLZL, and the interference templates are compared with the corresponding \SHERPA[3] predictions, and the observed shape differences are assigned as systematic uncertainties.
Recently available NNLO QCD and combined NNLO~QCD+NLO~EW predictions for polarized $ZZ$ production~\cite{Carrivale:2025mjy} are found to be consistent with the reweighting approach within uncertainties.
%
%
%
%
%
%
%
%
%


%
Systematic uncertainties arise from both experimental and theoretical sources.
For theoretical systematic uncertainties, the uncertainty due to the choice of renormalization and factorization scales in the theory calculation is estimated by varying those scales independently by a factor of two, excluding combinations where the variations differ by a factor of four.
The envelope of the effects from these variations is taken as an estimate of the impact of missing higher-order corrections, referred to as the QCD scale uncertainty.
The PDF-related uncertainty for each of the signal samples is estimated by using the PDF4LHC recommendation~\cite{Butterworth:2015oua}.
The parton showering and hadronization uncertainty is estimated for the signal by comparing the nominal \PYTHIA[8] parton showering with the alternative \HERWIG[7]~\cite{Bellm:2015jjp, Bahr:2008pv} algorithm.
Uncertainties from the NLO EW corrections for \qqtoZZ\ are estimated by taking the difference between the additive and multiplicative prescription in \MoCaNLO~\cite{Denner:2021csi}.
For the three polarized templates, the difference between the reweighted \MGNLO\ predictions and the \Sherpa[3] predictions are taken as an additional shape uncertainties.
These uncertainties are treated as uncorrelated between the different polarization states and can reach up to 11\%.
For the interference template, the statistical uncertainties resulting from the subtraction of \Sherpa[3] event weights, are considered as systematic uncertainties.
For the \ggtoZZ\ process, a conservative 60\% uncertainty is assigned to account for the missing higher-order corrections~\cite{Caola:2015psa}.
This uncertainty has a subdominant impact on the measurement. An alternative \ggtoZZ\ prediction by applying NLO QCD $k$-factors as a function of \mfl calculated at $\sqrt{s}=13$~TeV \cite{ggZZQCDCorr_1} gives consistent measurement results.
Experimental systematic uncertainties arise from reconstruction, identification, isolation, energy and momentum scale, resolution, and trigger efficiencies for electrons
and muons~\cite{MUON-2018-03,TRIG-2018-01,EGAM-2021-01,EGAM-2021-02}.
The overall uncertainty in the integrated luminosity is 1.9\%~\cite{ATL-DAPR-PUB-2025-001}.
Uncertainties in the estimate of the reducible and irreducible background processes are also considered and their impact is found to be negligible.


%
To determine the contribution of the \ZLZL\ production, a binned profile likelihood fit~\cite{Moneta:2010pm,RooFit,Cranmer:1456844} is performed using the measured BDT distribution with the TRExFitter framework~\cite{TRExFitter}.
The low-mass and the high-mass SRs are fitted simultaneously.
The normalization factors of the \ZLZL\ and \ZTZX\ templates relative to the SM expectations, referred to as the signal strengths \muZZLL\ and \muZZTX,
are free parameters in the fit, and they are extracted using a profile-likelihood-ratio test statistic~\cite{Cowan:2010js}.
This measurement assumes the SM prediction for the ratio between \ZTZT\ and \ZTZL\ production.
The interference term, which is proportional to $\sqrt{\muZZLL\muZZTX}$, is taken into account when building the probability model,
and the details can be found in the End Matter.
Systematic uncertainties are modeled as Gaussian-constrained nuisance parameters in the likelihood.

\begin{figure}[htb]
\centering
\subfloat[]{\includegraphics[width = 0.48 \textwidth]{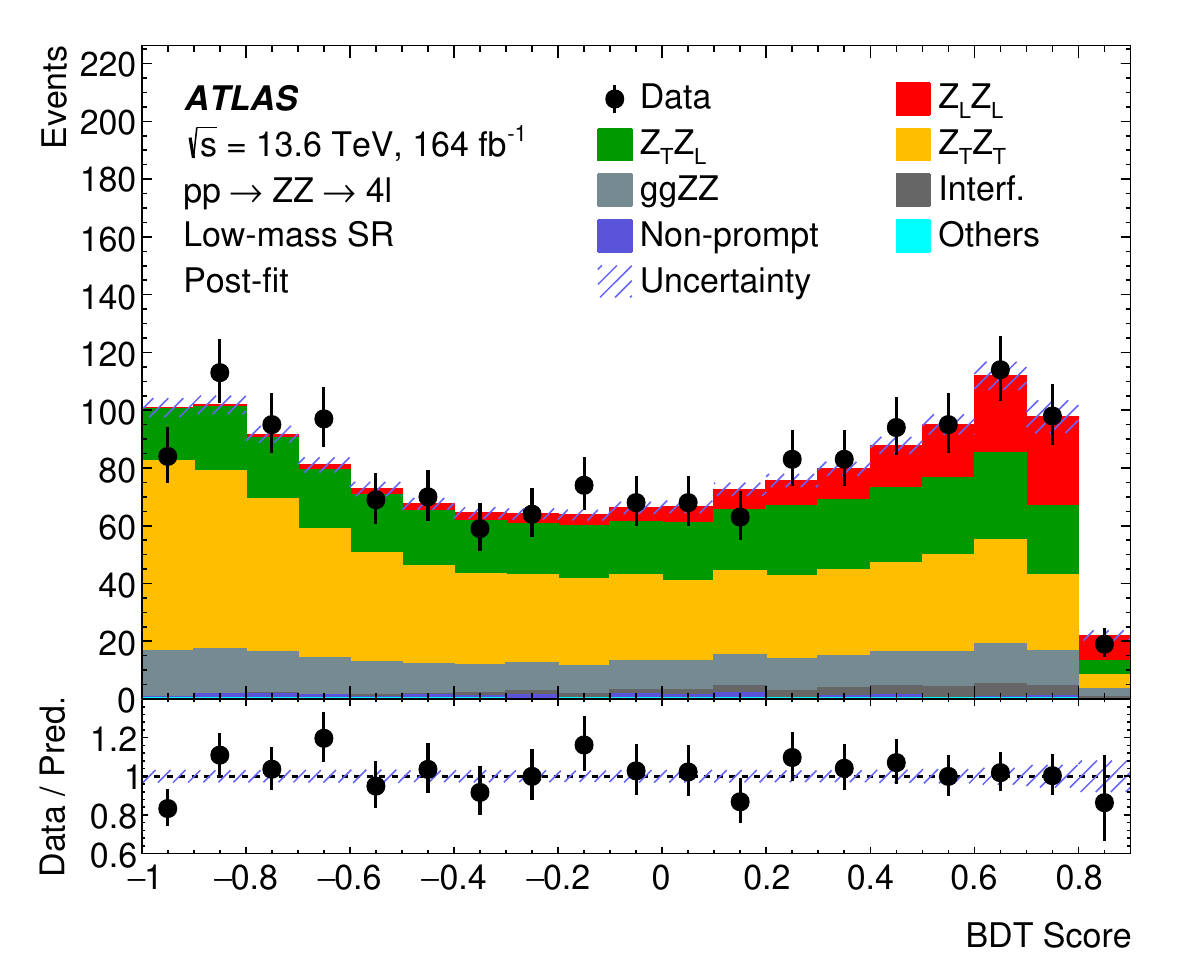}}
\subfloat[]{\includegraphics[width = 0.48 \textwidth]{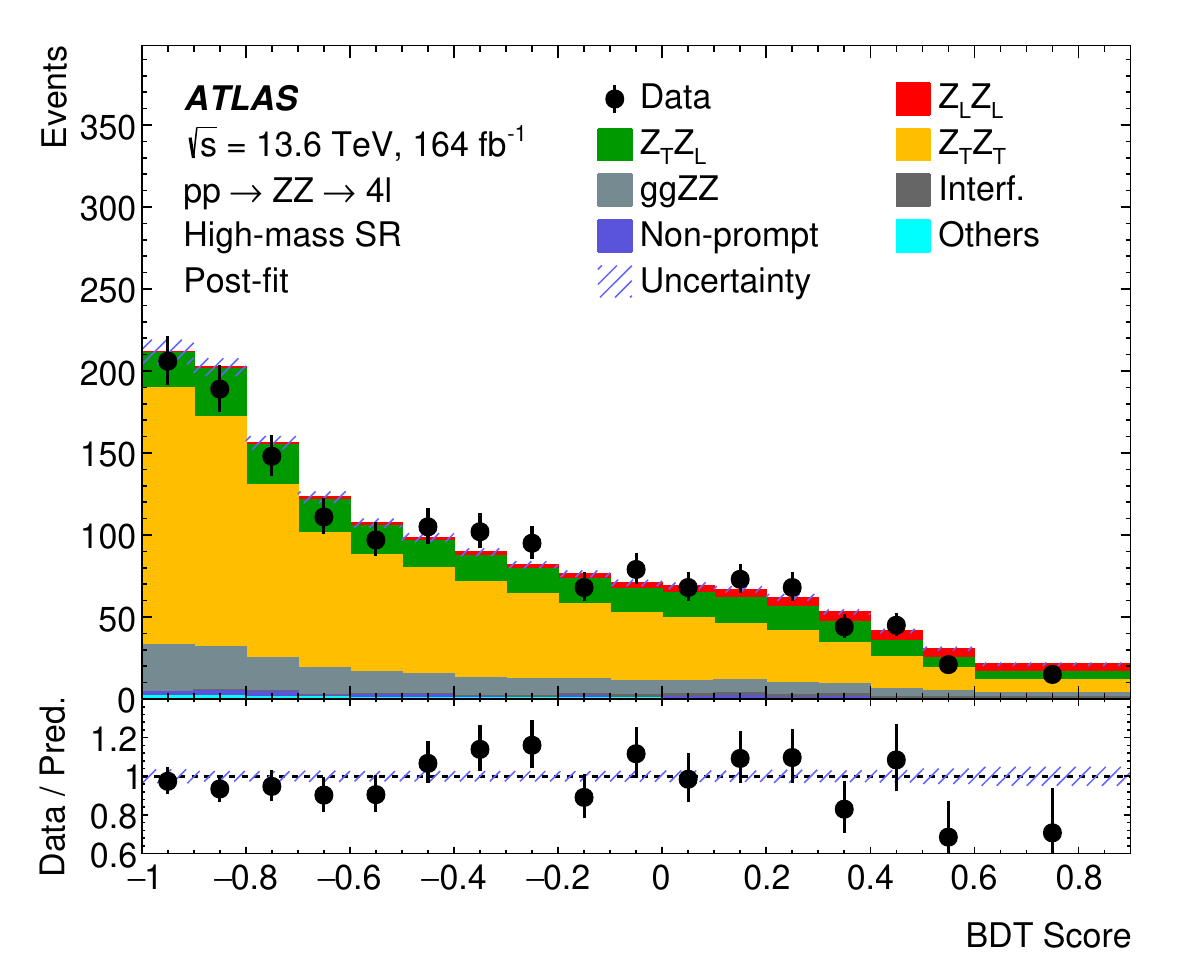}}
\caption{The BDT distribution of the data and the post-fit SM predictions in the (a) low-mass region and (b) high-mass region.
Results are obtained from the $(\muZZLL, \muZZTX)$ fit.
The `Others' category represents the contribution from $\ttbar Z$ and $VVZ$.
The lower panel shows the ratio of the data points to the post-fit total prediction.
The uncertainty bands include both of the statistical and systematic uncertainties as obtained by the fit.
}
\label{fig:pol_bdt_postfit}
\end{figure}

\begin{figure}[htb]
\centering
\subfloat[]{\includegraphics[width = 0.48 \textwidth]{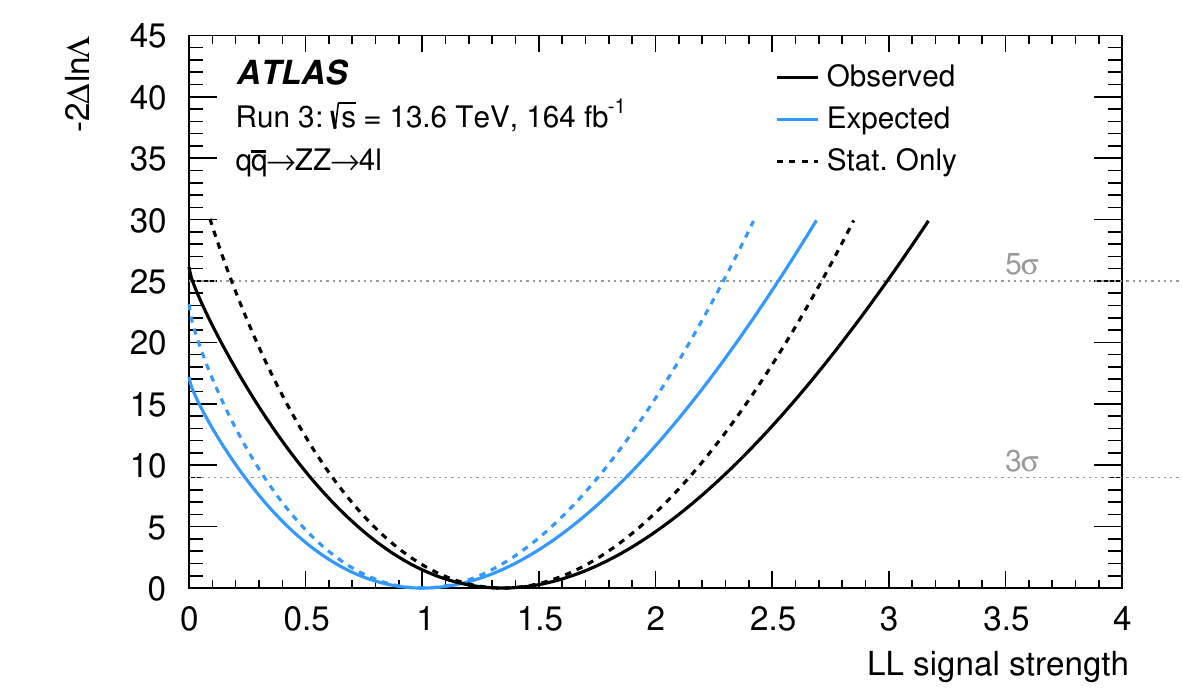}}
\subfloat[]{\includegraphics[width = 0.48 \textwidth]{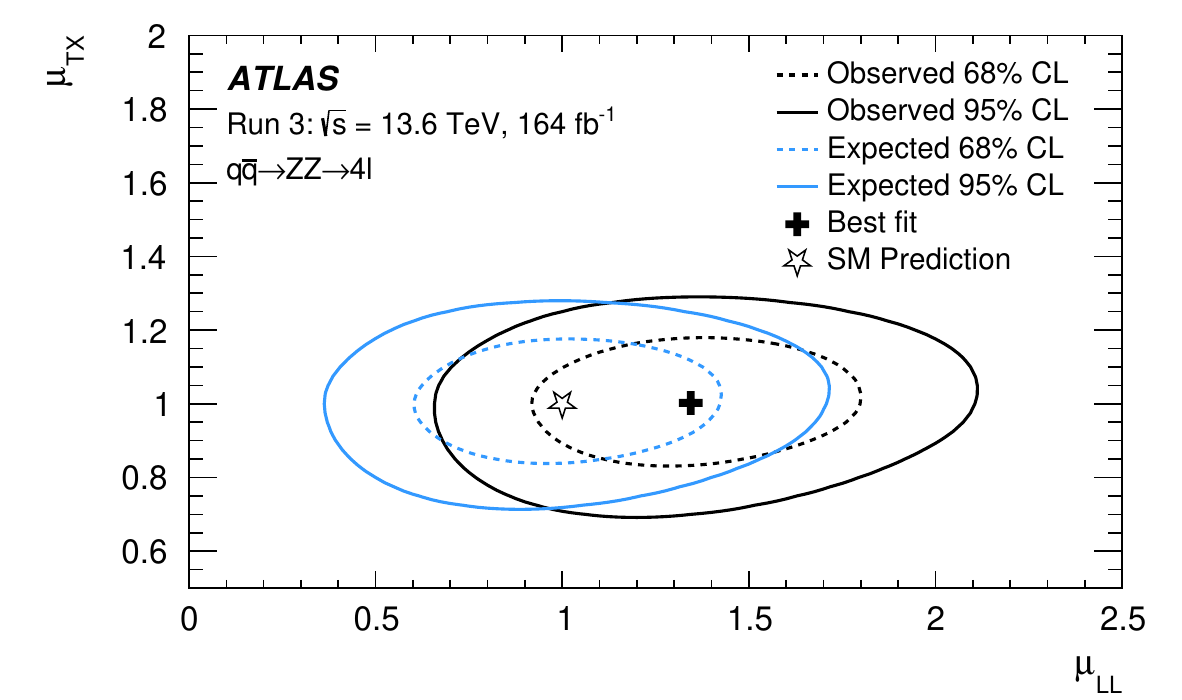}}
\caption{ (a) Negative profile log-likelihood ratio scans as a function of the $\muZZLL$ and (b) the two-dimensional contour for $(\muZZLL, \muZZTX)$
at the 68\% (dashed line) and 95\% (solid line) CL.
Black lines indicate the observed scans, while blue lines show the expected results.
The dashed horizontal lines in (a) indicate the thresholds corresponding to the $3~\sigma$ and $5~\sigma$ intervals, assuming the asymptotic approximation.
}
\label{fig:scan_Run3}
\end{figure}

Figure~\ref{fig:pol_bdt_postfit} shows the post-fit BDT distributions in the low-mass and high-mass SRs.
The corresponding post-fit yields are detailed in Table~\ref{tab:pol_yields_Run3}.
The \ZLZL\ signal strength is measured to be $\muZZLL = \valBkObsmuZZLL = \valObsmuZZLL$,
corresponding to a significance of $\sigObsmuZZLL~\sigma$.
The measured results are consistent with the SM expectation of $\muZZLL = \valExpmuZZLL$ and the expected significance of $\sigExpmuZZLL~\sigma$.
The normalization factor for the \ZTZX\ template is measured to be $\muZZTX = \valObsmuZZTX$, also consistent with the SM prediction.
The observed and expected negative log-likelihood ratio~($-2\Delta \ln \Lambda$) scans as a function of \muZZLL\ are shown in
Figure~\ref{fig:scan_Run3}(a).
The 2D contours for \muZZLL\ versus \muZZTX\ are also shown in Figure~\ref{fig:scan_Run3}(b).

\begin{table}[!htbp]
\caption{
Observed and post-fit numbers of events in the signal regions.
Results are obtained from the $(\muZZLL, \muZZTX)$ fit.
The `Others' category represents the contribution from $\ttbar Z$ and $VVZ$.
The uncertainties include both of the statistical and systematic contributions. The uncertainty in the total yield is
smaller than the quadrature sum of the contributions because of correlations resulting from the fit.
}
\label{tab:pol_yields_Run3}
\centering
\begin{tabular}{l r@{\,}l!{$\pm$}l@{\,}l r@{\,}l!{$\pm$}l@{\,}l}
\toprule
\toprule
Process & \multicolumn{4}{c}{Low-mass SR} & \multicolumn{4}{c}{High-mass SR} \\
\midrule
\ZLZL\             & 154& & 32& & 56& & 12 & \\
\ZTZL\             & 395& & 32& & 268& & 21 & \\
\ZTZT\             & 665& & 54& & 989& & 77 & \\
Interference       & 31.1& & 4.3& & 18.8& & 2.8 & \\
\ggtoZZ\            & 220& & 110& & 196& & 98 & \\
Non-prompt          & 15.3& & 5.6& & 19.2& & 6.5 & \\
Others              & 7.2& & 1.3& & 14.1& & 2.7 & \\
\midrule
Total            & 1483& & 36& & 1561& & 32 & \\
\midrule
Data             & \multicolumn{4}{c}{1510} & \multicolumn{4}{c}{1534} \\
\bottomrule
\bottomrule
\end{tabular}
\end{table}

A second fit is performed to extract the fraction of the \ZLZL\ production, with the inclusive $ZZ$ normalization factor, \muZZ,
and the \ZLZL\ fraction as free parameters of the fit.
The factor \muZZ\ is applied to the inclusive \qqtoZZ\ and EW $ZZjj$ production, among which the fraction of the \ZLZL\ production is defined as \fLL.
The inclusive \ZLZL\ fraction is measured to be $\fLL = \valObsfLL$, consistent with the SM prediction of $\valExpfLL$,
and the corresponding inclusive $ZZ$ normalization factor is found to be $\muZZ = 1.03 \pm 0.10$.
The \ZLZL\ fraction is also measured in the two \mfl\ regions simultaneously,
with $\fLLlowM = \valObsfLLlowM$ and $\fLLhighM = \valObsfLLhighM$,
consistent with the SM predictions of $\valExpfLLlowM$ and $\valExpfLLhighM$, respectively.
The lower measured \fLLhighM\ originates from a downward fluctuation of the data in the two most sensitive BDT bins, as shown in Figure~\ref{fig:pol_bdt_postfit}~(b).
The fiducial cross-section times branching fraction for the \ZLZL\ process is extracted in a region matching as closely as possible to the detector region.
The phase space definition requires two SFOC pairs of particle-level prompt leptons.
The particle-level leptons are \textit{dressed} by adding the four-momenta of nearby prompt photons within a small cone of $\Delta R < 0.1$.
The dressed electrons (muons) are required to satisfy $\pT > 7~(5)$~GeV and $|\eta| < 2.47 ~(2.5)$,
and the two highest-\pT leptons must have $\pT > 27$ and 20~GeV, respectively.
Leptons are required to be separated by $\Delta R > 0.05$, and any SFOC lepton pair is required to have an invariant mass $m_{\ell\ell} > 5$~GeV.
A quadruplet is constructed from the two SFOC pairs that minimize the sum of $|m_{\ell\ell} - \mZ|$.
The SFOC lepton pair with the larger $|y_{\ell\ell}|$ is defined as the leading $Z$ boson candidate.
On-shell requirements of the two $Z$ bosons are also applied: $|m_{\ell\ell} - \mZ | < 10$~GeV for each SFOC lepton pair and $m_{4\ell}>180$~GeV.
The contribution of electrons and muons from $\tau$-lepton decays is negligible in the fiducial region, and it is not included in the SM prediction.

The fiducial cross-section of the \ZLZL\ production is measured to be $\sigma^{\mathrm{obs.}}_{\ZLZL} = \xsecBkObsZZLL\,\mathrm{fb} = \xsecObsZZLL\,\mathrm{fb}$,
consistent with the SM prediction of $\sigma^{\mathrm{pred.}}_{\ZLZL} = \xsecExpZZLL$~fb.
The SM prediction includes NLO QCD and EW corrections for the \qqtoZZ\ process, calculated from \MoCaNLO,
and the LO prediction for the EW $ZZjj$ process.
The measurement is limited by statistical uncertainty of the data and the leading systematic uncertainty is from the theoretical modeling of the polarization templates.

\begin{figure}[htb]
\centering
\subfloat[]{\includegraphics[width=0.48\textwidth]{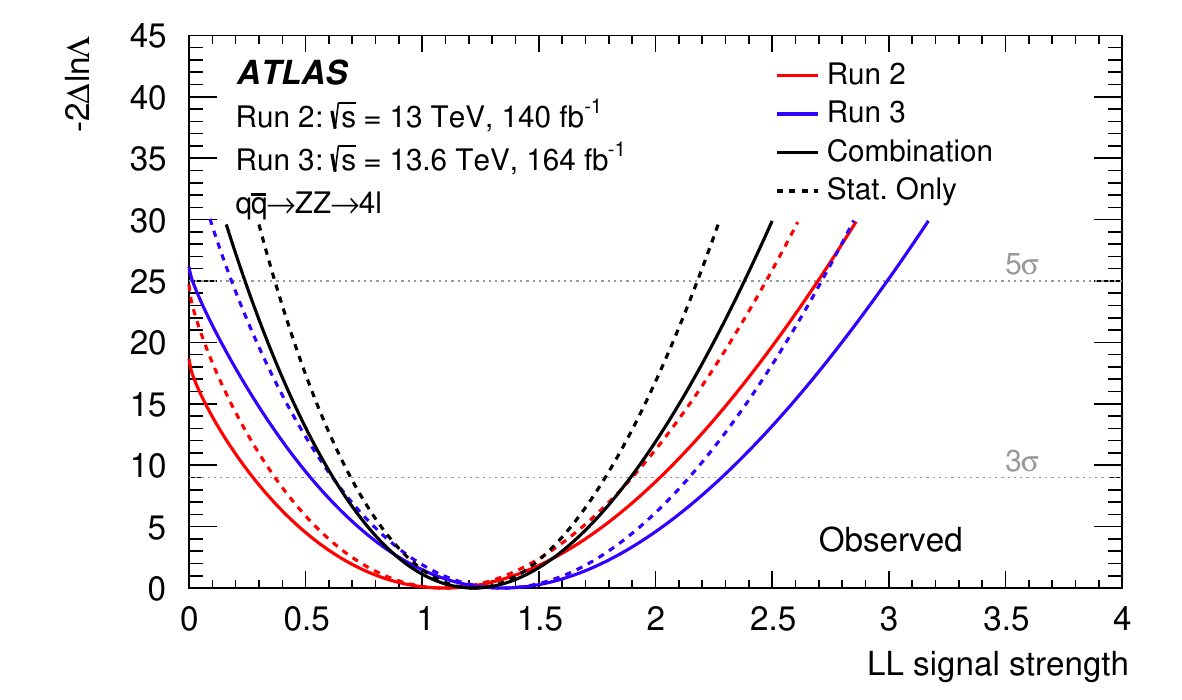}}
\subfloat[]{\includegraphics[width=0.48\textwidth]{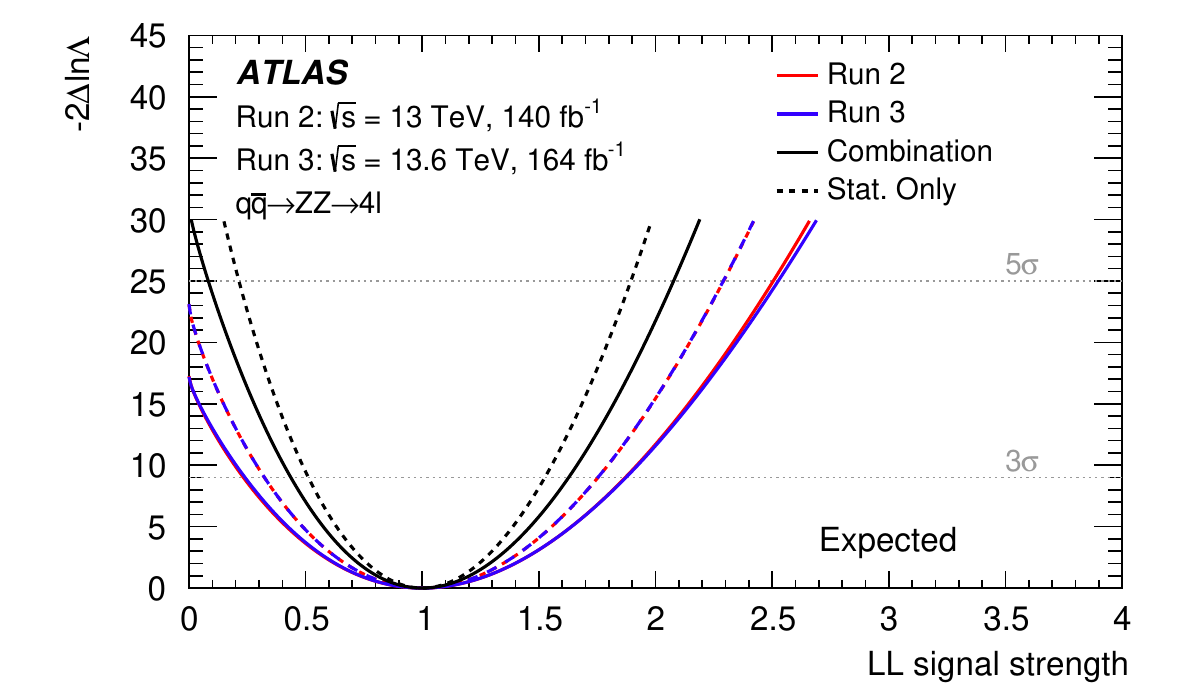}}
\caption{(a) Observed and (b) expected profile likelihood ratio as a function of the signal strength \muZZLL,
using the Run~2 and Run~3 data and their combination.
The dashed horizontal lines indicate the thresholds corresponding to the $3 \sigma$ and $5 \sigma$ intervals, assuming the asymptotic approximation.
}
\label{fig:comb}
\end{figure}

The \muZZLL\ measurement presented in this Letter is combined statistically with the Run~2 result~\cite{STDM-2021-05}.
Some aspects of the Run~2 analysis are revised to ensure consistency with the Run~3 methodology adopted in this article.
Specifically, the \ggtoZZ\ process is treated as a background, the polarization interference template is taken from \Sherpa[3],
and the theoretical modeling associated with the reweighting is taken to be the difference between the reweighted \MGNLO\ prediction and the corresponding \Sherpa[3] prediction.
The normalization of the interference term is updated to be scaled with $\sqrt{\muZZLL\muZZTX}$.
All other aspects of the Run~2 analysis remain unchanged.
The \ZLZL\ signal strength is measured to be $\muZZLL = \valBkObsmuZZLLRunTwoNew = \valObsmuZZLLRunTwoNew$.
The observed and expected significances are $\sigObsmuZZLLRunTwoNew$ and $\sigExpmuZZLLRunTwoNew~\sigma$, respectively.
For the combination of the Run~2 and Run~3 measurements, the combined likelihood is formed as the product of the individual analysis likelihoods, accounting for correlations among systematic uncertainties where applicable.
Different correlation schemes of systematic uncertainties are evaluated and the expected impact is found to be negligible.
Experimental uncertainties originating from the same source are taken as fully correlated.
Theory sources are also treated as fully correlated, except for those related to the reweighting of the polarized templates and PDFs, due to the different PDFs used.
Uncertainties associated with the non-prompt background estimate are uncorrelated.
The profile likelihood ratio as a function of \muZZLL\ for the combined fit is shown in Figure~\ref{fig:comb}.
The measured signal strength in the combined fit is $\muZZLL = \valObsmuZZLLComb$,
corresponding to a significance of $\sigObsmuZZLLComb~\sigma$, with an expected significance of $\sigExpmuZZLLComb~\sigma$.
The impact of each source of uncertainty is shown in Table~\ref{tab:table_ranking_groups_comb}.
The statistical uncertainty of the data is the dominant source of uncertainty, while the modeling of the polarization templates constitutes the largest systematic uncertainty.

\begin{table*}[htb!]
\centering
\small
\caption{
Impact of uncertainties in the measured \ZLZL\ signal strength \muZZLL\ from the combined fit.
Uncertainties are grouped by source type to illustrate their relative contributions.
}
\label{tab:table_ranking_groups_comb}
\begin{tabular}{lr}
\toprule
\toprule
Contribution & Relative uncertainty $[\%]$ \tabularnewline
Total & \ensuremath{17} \tabularnewline
\midrule Data statistical uncertainty & \ensuremath{15} \tabularnewline
Total systematic uncertainty & \ensuremath{8.1} \tabularnewline
\midrule
MC statistical uncertainty & \ensuremath{1.6} \tabularnewline
Theoretical systematic uncertainties & \ensuremath{} \tabularnewline
\hspace*{4mm} \qqtoZZ\ interference modeling & \ensuremath{0.9} \tabularnewline
\hspace*{4mm} \Sherpa[3] shape difference for \qqtoZZ & \ensuremath{4.4} \tabularnewline
\hspace*{4mm} QCD scale for \qqtoZZ & \ensuremath{0.5} \tabularnewline
\hspace*{4mm} PDF and $\alpha_s$ for \qqtoZZ & \ensuremath{1.9} \tabularnewline
\hspace*{4mm} Parton shower for \qqtoZZ & \ensuremath{1.6} \tabularnewline
\hspace*{4mm} NLO EW corrections for \qqtoZZ & \ensuremath{2.9} \tabularnewline
\hspace*{4mm} QCD scale for \ggtoZZ & \ensuremath{3.4} \tabularnewline
\hspace*{4mm} PDF and $\alpha_s$ for \ggtoZZ & \ensuremath{0.2} \tabularnewline
Experimental systematic uncertainties & \ensuremath{} \tabularnewline
\hspace*{4mm} Luminosity & \ensuremath{1.3} \tabularnewline
\hspace*{4mm} Muons & \ensuremath{0.5} \tabularnewline
\hspace*{4mm} Electrons & \ensuremath{3.7} \tabularnewline
\hspace*{4mm} Non-prompt background & \ensuremath{0.3} \tabularnewline
\hspace*{4mm} \Pileup modeling & \ensuremath{0.5} \tabularnewline
\hspace*{4mm} Triboson and $\ttbar Z$ normalizations & \ensuremath{0.2} \tabularnewline
\bottomrule
\bottomrule
\end{tabular}
\end{table*}


%
In summary, this Letter reports the polarization measurement of $Z$ bosons in the $ZZ\to \fourl$ channel,
using proton--proton collisions at the LHC collected with the ATLAS detector at $\sqrt{s}=13.6$~TeV with an integrated luminosity of 164~fb$^{-1}$.
The production of two simultaneously longitudinally polarized $Z$ bosons, excluding the \ggtoZZ\ process, is measured with observed and expected significances of
\sigObsmuZZLL\ and \sigExpmuZZLL\ standard deviations, respectively.
The inclusive \ZLZL\ fraction is measured for the first time and is $\fLL = \valObsfLL$, consistent with the SM prediction of $\valExpfLL$.
The production cross-section of \ZLZL\ events is measured in a fiducial phase space as \xsecObsZZLL~fb, consistent with the Standard Model prediction
of \xsecExpZZLL~fb with NLO QCD and EW corrections considered for the \qqtoZZ\ process.
With a statistical combination of the previous study using Run~2 data at $\sqrt{s}=13$~TeV with an integrated luminosity of 140~fb$^{-1}$,
the first observation of two longitudinally polarized $Z$ bosons is achieved,
with an observed and expected significance of \sigObsmuZZLLComb\ and \sigExpmuZZLLComb\ standard deviations, respectively.


%
%
%
%
%
\section*{Acknowledgements}
%

%

%
%

%
%

We thank CERN for the very successful operation of the LHC and its injectors, as well as the support staff at
CERN and at our institutions worldwide without whom ATLAS could not be operated efficiently.

The crucial computing support from all WLCG partners is acknowledged gratefully, in particular from CERN, the ATLAS Tier-1 facilities at TRIUMF/SFU (Canada), NDGF (Denmark, Norway, Sweden), CC-IN2P3 (France), KIT/GridKA (Germany), INFN-CNAF (Italy), NL-T1 (Netherlands), PIC (Spain), RAL (UK) and BNL (USA), the Tier-2 facilities worldwide and large non-WLCG resource providers. Major contributors of computing resources are listed in Ref.~\cite{ATL-SOFT-PUB-2026-001}.

We gratefully acknowledge the support of ANPCyT, Argentina; YerPhI, Armenia; ARC, Australia; BMWFW and FWF, Austria; ANAS, Azerbaijan; CNPq and FAPESP, Brazil; NSERC, NRC and CFI, Canada; CERN; ANID, Chile; CAS, MOST, NSFC and FRFCU, China; Minciencias, Colombia; MEYS CR, Czech Republic; DNRF and DNSRC, Denmark; IN2P3-CNRS and CEA-DRF/IRFU, France; SRNSFG, Georgia; BMFTR, HGF and MPG, Germany; GSRI, Greece; RGC and Hong Kong SAR, China; ICHEP and Academy of Sciences and Humanities, Israel; INFN, Italy; MEXT and JSPS, Japan; CNRST, Morocco; NWO, Netherlands; RCN, Norway; MNiSW, Poland; FCT, Portugal; MNE/IFA, Romania; MSTDI, Serbia; MSSR, Slovakia; ARIS and MVZI, Slovenia; DSI/NRF, South Africa; MICIU/AEI, Spain; SRC and Wallenberg Foundation, Sweden; SERI, SNSF and Cantons of Bern and Geneva, Switzerland; NSTC, Taipei; TENMAK, T\"urkiye; STFC/UKRI, United Kingdom; DOE and NSF, United States of America.

Individual groups and members have received support from BCKDF, CANARIE, CRC and DRAC, Canada; CERN-CZ, FORTE and PRIMUS, Czech Republic; COST, ERC, ERDF, Horizon 2020 and Marie Sk{\l}odowska-Curie Actions, European Union; Investissements d'Avenir Labex, Investissements d'Avenir Idex and ANR, France; DFG and AvH Foundation, Germany; Herakleitos, Thales and Aristeia programmes co-financed by EU-ESF and the Greek NSRF, Greece; BSF-NSF and MINERVA, Israel; NCN and NAWA, Poland; La Caixa Banking Foundation, CERCA and AGAUR programs from Generalitat de Catalunya and PROMETEO and GenT Programmes Generalitat Valenciana, Spain; G\"{o}ran Gustafssons Stiftelse, Sweden; The Royal Society and Leverhulme Trust, United Kingdom; Eric and Wendy Schmidt Fund for Strategic Innovation, United States of America.

In addition, individual members wish to acknowledge support from Chile: Agencia Nacional de Investigaci\'on y Desarrollo (ANID FONDECYT reg. 1230987, FONDECYT 1230812, FONDECYT 1240864, Fondecyt 3240661, Fondecyt Regular 1240721); China: Fundamental Research Funds for the Central Universities (010-63263105), Chinese Ministry of Science and Technology (MOST-2023YFA1605700, MOST-2023YFA1609300), National Natural Science Foundation of China (NSFC 12275265, NSFC-W2543005); Czech Republic: Czech Science Foundation (GACR - 24-11373S), Ministry of Education Youth and Sports (ERC-CZ-LL2327, FORTE CZ.02.01.01/00/22\_008/0004632); EU: H2020 European Research Council (ERC - 101002463); European Union: European Research Council (BARD No. 101116429, ERC - 101219398, ERC 101089007), European Regional Development Fund (HE COFUND GA No.101081355, ERDF), Marie Sklodowska-Curie Actions (GAP-101168829); France: Agence Nationale de la Recherche (ANR-22-EDIR-0002, ANR-24-CE31-0504-01); Germany: Deutsche Forschungsgemeinschaft (DFG - 469666862); China: Research Grants Council (GRF); Italy: Istituto Nazionale di Fisica Nucleare (LHC-MIUR - 28003/2025), Ministero dell'Università e della Ricerca (NextGenEU I53D23000820006 M4C2.1.1, SOE2024\_0000023); Japan: Japan Society for the Promotion of Science (JSPS KAKENHI  JP25H0063, JSPS KAKENHI JP22H04944, JSPS KAKENHI JP24K23939, JSPS KAKENHI JP24KK0251, JSPS KAKENHI JP25H00650, JSPS KAKENHI JP25H01291, JSPS KAKENHI JP25K01011, JSPS KAKENHI JP25K01023, JSPS KAKENHI JP25KK0047); Poland: Polish National Science Centre (NCN 2021/42/E/ST2/00350, NCN OPUS 2023/51/B/ST2/02507, NCN OPUS nr 2022/47/B/ST2/03059, NCN UMO-2019/34/E/ST2/00393, UMO-2022/47/O/ST2/00148, UMO-2023/49/B/ST2/04085, UMO-2023/51/B/ST2/00920, UMO-2024/53/N/ST2/00869); Spain: Agència de Gestió d'Ajuts Universitaris i de Recerca. (AGAUR - 2023 BP 00141), Ministry of Science and Innovation (RYC2019-028510-I, RYC2020-030254-I, RYC2021-031273-I, RYC2022-038164-I), Ministerio de Ciencia, Innovación y Universidades/Agencia Estatal de Investigaci\'on (EU NextGenerationEU (PRTR-C17.I1), PID2022-142604OB-C22); Sweden: Carl Trygger Foundation (Carl Trygger Foundation CTS 22:2312), Swedish Research Council (Swedish Research Council 2023-04654, VR 2021-03651, VR 2022-03845, VR 2022-04683, VR 2023-03403, VR 2024-05451, VR 2025-05940), Knut and Alice Wallenberg Foundation (KAW 2023.0366); United Kingdom: The Binks Trust, Royal Society (NIF-R1-231091); United States of America: U.S. Department of Energy (ECA DE-AC02-76SF00515), John Templeton Foundation (John Templeton Foundation 63206), Neubauer Family Foundation.

%
%


%
\clearpage
\appendix
\part*{End Matter}
\addcontentsline{toc}{part}{Appendix}

%
When building the probability model of the likelihood for the \muZZLL\ signal strength measurement, the expected number of events in a given signal region,
expressed as a function of the BDT score, $x$, is given by
\begin{equation}
N(\muZZLL, \muZZTX, x) = \muZZLL\HZZLL(x)+\muZZTX\HZZTX(x)+\sqrt{\muZZLL\muZZTX}\HZZint(x)+\HZZBkg(x),
\label{eq:muLL_float_int}
\end{equation}
where $\HZZLL(x), \HZZTX(x), \HZZint(x)$ and $\HZZBkg(x)$ represent the expected yields of the \ZLZL\ polarization state, \ZTZX\ polarization state,
the interference and the other background process, respectively.

For the \fLL\ measurement, the expected number of events in a given signal region is given by
\begin{equation}
N(\fLL, \muZZ, x) = \muZZ N_{\mathrm{incl}}^{\mathrm{SM}}(\fLL \hZZLL(x)+ \fTX\hZZTX(x) + \kappa\sqrt{\fLL\fTX}\hZZint(x))+\HZZBkg(x),
\label{eq:ffLL_float_int}
\end{equation}
where $\hZZLL(x), \hZZTX(x)$ and $\hZZint(x)$ represent the normalized template of the \ZLZL\ polarization state, \ZTZX\ polarization state
and the interference, respectively; $N_{\mathrm{incl}}^{\mathrm{SM}}$ is the expected total yield of the inclusive \qqtoZZ\ and EW $ZZjj$ production.
The fraction of the \ZTZX\ production, $\fTX$, is not an independent parameter and must satisfy the unity requirement,
after summing over all bins of $x$,
\begin{equation}
\fLL + \fTX + \kappa\sqrt{\fLL\fTX} =1,
\label{eq:ffLL_kappa}
\end{equation}
where $\kappa$ is a constant predicted by the SM,
$\kappa = \frac{N_{\mathrm{int}}^{\mathrm{SM}}}{\sqrt{N_{\mathrm{LL}}^{\mathrm{SM}}N_{\mathrm{TX}}^{\mathrm{SM}}}}$,
with $N_{\mathrm{LL}}^{\mathrm{SM}}, N_{\mathrm{TX}}^{\mathrm{SM}}$ and $N_{\mathrm{int}}^{\mathrm{SM}}$ being the total expected yield of the
the \ZLZL\ polarization state, \ZTZX\ polarization state, and the interference, respectively, in a given SR.
For the \ZLZL\ fraction measurement in two \mfl\ regions, the above parameterizations are performed separately for each \mfl\ region.
The physical, non-negative solution for \fTX\ of Equation~\ref{eq:ffLL_kappa} is used.

Alternative fit results for \muZZLL\ and $\fLL$ are also presented using a different treatment of the interference-term normalization in the likelihood.
In this approach, the normalization of the interference term is decoupled from \muZZLL\ and fixed to SM prediction within theory uncertainties,
providing a more conservative treatment. The corresponding event yield parameterization is given by
\begin{align}
N(\muZZLL, \muZZTX, x) & = \muZZLL\HZZLL(x)+\muZZTX\HZZTX(x)+\HZZint(x)+\HZZBkg(x), \label{eq:muLL_1} \\
N(\fLL, \muZZ, x) & = \muZZ N_{\mathrm{incl}}^{\mathrm{SM}}(\fLL \hZZLL(x)+ (1-\fLL)(\hZZTX(x)+\hZZint(x)))+\HZZBkg(x). \label{eq:ffLL_1}
\end{align}
The resulting measurement results are summarized in Table~\ref{tab:muLL_res_comp} and Table~\ref{tab:fLL_res_comp}.
For both \muZZLL\ and \fLL\ measurements, the corresponding central values are close to the nominal fit results, with the total uncertainties increased by $6\sim8$~\% as expected.
\begin{table}[ht]
\centering
\small
\caption{Comparison of the \muZZLL\ measurement results with two different treatments of the normalization of the interference term, either scaled with $\sqrt{\muZZLL\muZZTX}$ or fixed to the SM prediction.\label{tab:muLL_res_comp}}
\begin{tabular}{l c c c c}
\toprule
& \multicolumn{2}{c}{\muZZLL}  & \multicolumn{2}{c}{Significance~($\sigma$)} \\
\cmidrule{2-5}
& Observed & Expected & Observed & Expected \\
\midrule
\multicolumn{5}{c}{\textbf{Scaled normalization for the interference term}} \\
\midrule
Run~3      &  $\valObsmuZZLL$ & $\valExpmuZZLL$ & $\sigObsmuZZLL$ & $\sigExpmuZZLL$ \\
Run~2      &  $\valObsmuZZLLRunTwoNew$ & $1.00 \pm 0.27$ & $\sigObsmuZZLLRunTwoNew$ & $\sigExpmuZZLLRunTwoNew$ \\
Combined   &  $\valObsmuZZLLComb$ & $1.00 \pm 0.20$ & $\sigObsmuZZLLComb$ & $\sigExpmuZZLLComb$ \\
\midrule
\multicolumn{5}{c}{\textbf{Fixed normalization for the interference term}} \\
\midrule
Run~3      &  $1.35 \pm 0.31$ & $1.00 \pm 0.30$ & $4.5$ & $3.6$ \\
Run~2      &  $1.06 \pm 0.32$ & $1.00 \pm 0.30$ & $3.5$ & $3.5$ \\
Combined   &  $1.21 \pm 0.23$ & $1.00 \pm 0.22$ & $5.5$ & $4.8$ \\
\bottomrule
\end{tabular}
\end{table}

\begin{table}[ht]
\centering
\small
\caption{Comparison of the \fLL\ measurement results with two different treatments of the normalization of the interference term, either scaled with $\sqrt{\muZZLL\muZZTX}$ or fixed to the SM prediction.\label{tab:fLL_res_comp}}
\begin{tabular}{l p{4cm} p{4cm} c}
\toprule
& \multicolumn{2}{c}{Observed} & \multirow{2}{*}{Expected} \\
& Scaled normalization for the interference term & Fixed normalization for the interference term & \\
\midrule
\fLL        &  $\valObsfLL$      &  $0.080 \pm 0.018$ & $\valExpfLL$ \\ [5pt]
\fLLlowM    &  $\valObsfLLlowM$  &  $0.110 \pm 0.030$ & $\valExpfLLlowM$ \\ [6pt]
\fLLhighM   &  $\valObsfLLhighM$ &  $0.009 \pm 0.021$ & $\valExpfLLhighM$ \\[2pt]
\bottomrule
\end{tabular}
\end{table}


%

%
%

%
%
%
\clearpage
\printbibliography
\clearpage
\input{atlas_authlist}
%

%
%
\clearpage

\end{document}

%% file: atlas_authlist.tex
 
\begin{flushleft}
\hypersetup{urlcolor=black}
{\Large The ATLAS Collaboration}

\bigskip

\AtlasOrcid[0000-0002-6665-4934]{G.~Aad}$^\textrm{\scriptsize 102}$,
\AtlasOrcid[0000-0001-7616-1554]{E.~Aakvaag}$^\textrm{\scriptsize 17}$,
\AtlasOrcid[0000-0002-5888-2734]{B.~Abbott}$^\textrm{\scriptsize 121}$,
\AtlasOrcid[0000-0002-0287-5869]{S.~Abdelhameed}$^\textrm{\scriptsize 83b}$,
\AtlasOrcid[0000-0002-1002-1652]{K.~Abeling}$^\textrm{\scriptsize 54}$,
\AtlasOrcid[0000-0001-5763-2760]{N.J.~Abicht}$^\textrm{\scriptsize 48}$,
\AtlasOrcid[0000-0002-8496-9294]{S.H.~Abidi}$^\textrm{\scriptsize 30}$,
\AtlasOrcid[0009-0003-6578-220X]{M.~Aboelela}$^\textrm{\scriptsize 44}$,
\AtlasOrcid[0000-0002-9987-2292]{A.~Aboulhorma}$^\textrm{\scriptsize 36e}$,
\AtlasOrcid[0000-0001-5329-6640]{H.~Abramowicz}$^\textrm{\scriptsize 154}$
% ...
\bigskip
\\

% ...
$^{17}$Department for Physics and Technology, University of Bergen, Bergen; Norway.\\
% ...
$^{30}$Physics Department, Brookhaven National Laboratory, Upton NY; United States of America.\\
% ...
$^{36}$$^{(a)}$Facult\'e des Sciences Ain Chock, Universit\'e Hassan II de Casablanca;$^{(b)}$Facult\'{e} des Sciences, Universit\'{e} Ibn-Tofail, K\'{e}nitra;$^{(c)}$Facult\'e des Sciences Semlalia, Universit\'e Cadi Ayyad, LPHEA-Marrakech;$^{(d)}$LPMR, Facult\'e des Sciences, Universit\'e Mohamed Premier, Oujda;$^{(e)}$Facult\'e des sciences, Universit\'e Mohammed V, Rabat;$^{(f)}$Institute of Applied Physics, Mohammed VI Polytechnic University, Ben Guerir; Morocco.\\
% ...
$^{44}$Physics Department, Southern Methodist University, Dallas TX; United States of America.\\
% ...
$^{48}$Fakult\"{a}t Physik, Technische Universit{\"a}t Dortmund, Dortmund; Germany.\\
% ...
$^{54}$II. Physikalisches Institut, Georg-August-Universit\"{a}t G\"ottingen, G\"ottingen; Germany.\\
% ...
$^{83}$$^{(a)}$Khalifa University of Science and Technology, Abu Dhabi;$^{(b)}$New York University Abu Dhabi, Abu Dhabi;$^{(c)}$United Arab Emirates University, Al Ain;$^{(d)}$University of Sharjah, Sharjah; United Arab Emirates.\\
% ...
$^{102}$CPPM, Aix-Marseille Universit\'e, CNRS/IN2P3, Marseille; France.\\
% ...
$^{121}$Homer L. Dodge Department of Physics and Astronomy, University of Oklahoma, Norman OK; United States of America.\\
% ...
$^{154}$Raymond and Beverly Sackler School of Physics and Astronomy, Tel Aviv University, Tel Aviv; Israel.\\
% ...
$^{a}$ Also at Affiliated with an institute formerly covered by a cooperation agreement with CERN.\\
$^{b}$ Also at An-Najah National University, Nablus; Palestine.\\
$^{c}$ Also at Borough of Manhattan Community College, City University of New York, New York NY; United States of America.\\
$^{d}$ Also at Center for Interdisciplinary Research and Innovation (CIRI-AUTH), Thessaloniki; Greece.\\
$^{e}$ Also at Centre of Physics of the Universities of Minho and Porto (CF-UM-UP); Portugal.\\
$^{f}$ Also at CERN, Geneva; Switzerland.\\
% ...

\end{flushleft}
